\documentclass[conference]{IEEEtran}

\usepackage{atbegshi}
\AtBeginDocument{\AtBeginShipoutNext{\AtBeginShipoutDiscard}}
\usepackage[]{algorithm2e}
\usepackage{mathrsfs}
\usepackage{multirow}
\usepackage{enumerate}
\usepackage{comment}
\usepackage{tabularx}
\usepackage{amsmath}
\usepackage{breqn}
\usepackage{paralist}
\usepackage{float}
\usepackage{amsfonts}
\usepackage{lettrine}
\usepackage{balance}
\usepackage{mathtools}
\usepackage{amssymb}
\usepackage[font=small,skip=0pt]{caption}
\usepackage[dvipsnames]{xcolor}
\usepackage{graphicx}
\usepackage{gensymb}
\usepackage{lipsum} 
\usepackage{array}
\usepackage{soul}
\usepackage[short]{optidef}
\IEEEoverridecommandlockouts
\usepackage[utf8]{inputenc}
\usepackage[noadjust]{cite}
\usepackage[caption=false,font=normalsize,labelfont=sf,textfont=sf]{subfig}
\usepackage{textcomp}
\usepackage[export]{adjustbox}
\usepackage{url}
\usepackage[usestackEOL]{stackengine}

\usepackage{makecell}

\DeclareCaptionLabelSeparator{periodspace}{.\quad}
\begin{document}

\title{Automated Data-Driven Model Extraction and Validation of Inverter Dynamics with Grid Support Function}

\author{Sunil Subedi$^{\dagger}$,~\IEEEmembership{Student Member,~IEEE}, Bidur~Poudel,~\IEEEmembership{Student Member,~IEEE},\\ 
Pooja~Aslami,~\IEEEmembership{Student Member,~IEEE}, Robert~Fourney,~\IEEEmembership{Member,~IEEE},
Hossein~Moradi~Rekabdarkolaee,\\
Reinaldo~Tonkoski,~\IEEEmembership{Senior Member,~IEEE}, and Timothy~M.~Hansen,~\IEEEmembership{Senior Member,~IEEE}}%
\thanks{This work is supported by the U.S. Department of Energy Office of Science, Office of Electricity Microgrid R\&D Program, and Office of Energy Efficiency and Renewable Energy Solar Energy Technology Office under the EPSCoR grant number DE-SC0020281.}%
\thanks{S.~Subedi, B.~Poudel, P.~Aslami, R.~Fourney, and T.~M.~Hansen are with the Department of Electrical Engineering and Computer Science, and H.~M.~Rekabdarkolaee is with the Mathematics and Statistics Department, South Dakota State University, Brookings, South Dakota, 57007 USA (e-mails: sunil.subedi@jacks.sdstate.edu, bidur.poudel@jacks.sdstate.edu, pooja.aslami@jacks.sdstate.edu, robert.fourney@sdstate.edu, hossein.moradirekabdarkolaee@sdstate.edu, timothy.hansen@sdstate.edu).}%
\thanks{R.~Tonkoski is with the Electrical \& Computer Engineering Department, University of Maine, Orono, Maine, USA (e-mail: reinaldo.tonkoski@maine.edu).}%

\maketitle

\begin{abstract}
This research focuses on the evolving dynamics of the power grid, where traditional synchronous generators are being replaced by non-synchronous power electronic converter (PEC)-interfaced renewable energy sources. The non-linear dynamics must be accurately modeled to ensure the stability of future converter-dominated power systems (CDPS). However, obtaining comprehensive dynamic models becomes more complex and computationally intensive as the system grows. This study proposes a scalable and automated data-driven partitioned modeling framework for CDPS dynamics. The method constructs reduced-ordered dynamic linear transfer function models using input-output measurements from a PEC switching model. Validation experiments were conducted on single-house and multi-house scenarios, demonstrating high accuracy (over 97\%) and significant computational speed improvements (6.5 times faster) compared to comprehensive models. This framework and modeling approach offers valuable insights for efficient analysis of power system dynamics, aiding in planning, operation, and dispatch.

\end{abstract}

\maketitle

\section*{Nomenclature}
\begin{IEEEdescription}
\item[BIC]  ~~~~Bayes information criterion
\item[CDPS] ~~~~Converter dominated power systems
\item[AICc] ~~~~Corrected Akaike's information criterion
\item[DERs]    ~~~~Distributed  energy  resources 
\item[GSFs]   ~~~~Grid support functions
\item[IBRs]   ~~~~Inverter-based resources
\item[LTI]    ~~~~Linear time-invariant
\item[NRMSE]  ~~~~Normalized root mean squared error
\item[PECs] ~~~~Power electronic converters
\item[PCC]   ~~~~Point of common coupling
\item[PLL]   ~~~~Phase-locked loop
\item[PI] ~~~~Proportional-integral
\item[PV]    ~~~~Photovoltaic
\item[QSG]   ~~~~Quadrature signal generation
\item[RODLTF] ~~~~~Reduced-ordered dynamic linear transfer function
\item[RESs]  ~~~~Renewable energy sources
\item[SOGI] ~~~~Second-order generalized integral
\item[SysId] ~~~~System identification
\end{IEEEdescription}

\section{Introduction}

The extensive usage of power electronic converters (PECs) and distributed energy resources (DERs) has resulted in a paradigm shift from a historically passive to a more dynamic power system~\cite{ieee2021}. When DERs meet current and future energy demands, models that accurately describe the interplay between the grid and the PECs are essential. These inverter-based resources (IBRs) feature quicker-switching processes than traditional power systems, resulting in faster and more complex system dynamics. The impact of these dynamics has primarily been neglected because of the low fraction of IBRs and the converters' passive function in power system voltage and frequency management, and large synchronous generators drove traditional power system dynamics with well-defined models~\cite{kundur1994power}. However, the dynamics of the modern power system must be investigated in the \textit{converter-dominated power systems} (CDPS) as they affect controller design, optimization actions, supervisory actions, fault detection phenomena, component diagnostics, and instability prediction of complex systems. Reliable models are needed to investigate power system dynamics with IBR generation.

As per the IEEE 1547-2018 standard for DERs, along with the advancement in other grid codes, PECs equipped with several grid support functions (GSFs), such as Volt-Watt, Volt-VAr, frequency-Watt, and voltage/frequency ride through, are responsible for providing several voltage and frequency ancillary services. Depending on the manufacturer, these services are implemented in various ways, even under identical operating conditions~\cite{ieee1547}. Although smart PECs with such functions are intended to aid the integration of more IBRs to regulate grid voltage and frequency, maintain grid stability, and increase grid dependability, their specific design and control actions are proprietary, and topologies are unknown. These black-box factors can lead to errors in power system modeling, simulation, and PEC dynamics, resulting in inaccurate results and analysis. In addition, depending on the activation of states, there is a variation in power system dynamics (``state'' in the context of the research means the operation of the converter when it is providing one of the ancillary services). This reveals that the dynamics of PEC depend on various factors such as manufacturers, physical topology, voltage/current control loops, phase-locked loop (PLL) implementation, and different GSF standards~\cite{pll_sunil_2022}. Detailed switching models based on the manufacturers' most accurate PEC models, if they exist, are computationally intractable for real-time control. There is a need to design data-driven PEC models with GSFs in CDPS --- that does not depend on knowledge of the control structure nor physical topology --- at the device level for successful system planning, operation, and dispatch while avoiding the system dynamics challenges mentioned earlier.

The literature has several modeling methodologies aiming at accurately representing PEC dynamics while accelerating  simulations to overcome computational issues~\cite{Chinmay2021,sunil2020}. Switched models can be very accurate for representing the non-linear dynamics due to switching actions, control topology, and GSFs~\cite{Chinmay2021}. For the dynamic study of CDPS, accurate non-linear modeling of the converter's dynamics while providing multiple ancillary services is necessary; however, switched models are computationally demanding when the number of PECs increases~\cite{switchmodel2014}. Averaged linear models reduce computational costs by disregarding the effect of switching dynamics; however, they are less accurate in representing all of the process dynamics~\cite{average_2001}. Dynamic phasor models are more precise than average models, although they are technically more sophisticated~\cite{Raja2013}. On the other hand, black-box or data-driven system identification (SysId) techniques create PECs models from experimental data with little or no knowledge of the internal system physics (internal parameters, topology/control structure)~\cite{pico2018,nischal2020,Valdivia}, which is the approach taken in this work. 


During the training of black-box models, the selection of operating limits determines the number of the reduced-order dynamic linear transfer function (RODLTF) models; however, in the literature, the ranges are randomly selected. Moreover, PECs perturbation with appropriately designed probing signals is critical in data-driven modeling because they significantly impact the collected data and allow for accurate system characteristics estimation and identification of RODLTFs to represent system dynamics~\cite{Manisha2020}. Although linear dynamics are provided in correctly identified power systems using sinusoidal and exponential chirp probing signals, the impact on GSFs PECs when developing linear models of power systems has not been investigated~\cite{javier2021}. Furthermore, most methods proposed in the literature for the performance evaluation stage of derived models are manual. They are customized to a single system, making them time-consuming and limiting the feedback loop. Although PECs with advanced GSFs and probing signals designs play an essential role in the power system planning, operation, and dispatch, there have been no recommendations for automated data-driven simulation frameworks for IBRs in the literature. The proposed simulation framework is developed to derive RODLTF models with less computational power requirement to facilitate automated testing and validation.

 
Additionally, the predominance of non-linearities in the PEC dynamics leads to sophisticated dynamic models across many operating areas, especially with the addition of GSFs. In most of the previous data-driven modeling work, the SysId has been utilized to characterize the operating mode via single RODLTFs; however, one linearized model cannot accurately characterize the PEC dynamics across the entire operational region. Thus, based on the operating states looking at the converters' input-output characteristics, we design a combined model by splitting the active modes into piecewise linear portions, called ``partitioned models'' linearized around an operating point, and then aggregating them to derive the non-linear dynamics of PECs~\cite{sunil2021}. In this paper, we develop the automated data-driven partitioned modeling and validation simulation framework utilizing proven probing signals for extracting the detailed dynamics of PECs with GSFs. As a result, the developed simulation framework is computationally faster, more flexible, and more scalable. 

In our previous work~\cite{sunil2021}, the partitioned modeling approach was utilized as a proof-of-concept for extracting the dynamics of Volt-VAr GSF of a single PEC while demonstrating the computational feasibility of implementing this approach. This paper further expands the partitioned modeling algorithm to find the optimal number of RODLTFs using a binary search algorithm and information criteria. The method is then developed into a 12-house benchmark system with a significantly expanded results and analysis section. The main contributions of this paper are:

\begin{enumerate}

    
    \item Demonstrated the proposed partitioned modeling approach for extracting PV PEC with GSF dynamics using RODLTF models. 
    
    \item Designed an automated, scalable, and faster simulation framework for CDPS. 
    
    \item Developed the modified 12-house radial distribution system with dynamic loads and GSFs PEC. The proposed framework is validated under the Volt-VAr GSF, and the scalability is tested using a modified test system.
\end{enumerate}

The remainder of this paper is organized as follows. Section~\ref{Dynamic_modeling} introduces a black-box modeling representing the dynamics of a PEC with GSF, Section~\ref{partitioned_model} then introduces the concepts of the partitioned modeling approach and the speedup technique. Next, the single-house case study for developing partitioned models of a PEC in a co-simulation framework is described in Section~\ref{casestudy_singlehouse} followed by the case study implemented in a Canadian suburban distribution feeder, and detailed is described in Section~\ref{casestudy_twelvehouse}. Finally, Section~\ref {Results_and_analysis} describes the results and analysis, and the paper is concluded in Section~\ref{conclusion}.

\section{Dynamic Modeling of PECs with GSFs}\label{Dynamic_modeling}

This section discusses PEC dynamics in detail, followed by PEC black-box modeling with GSFs in Subsection~\ref{SysId}. Next, the probing signal design is mentioned in Subsection~\ref{Prob_sig}, and the metrics for the quantitative assessment are discussed in Subsection~\ref{Metrics}.

\subsection{PEC Dynamics Description}\label{PECdyna}

\begin{figure}[h!]
    \centering
    \includegraphics[width=0.9\columnwidth]{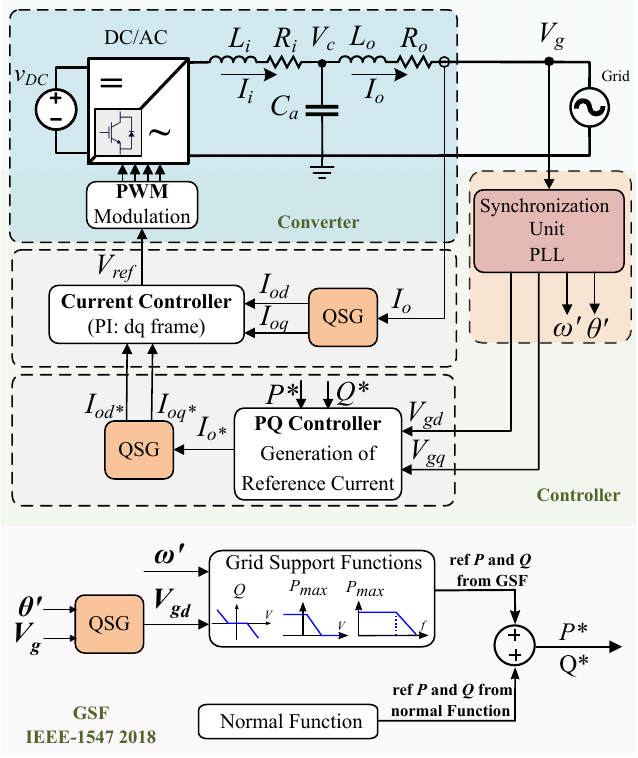}
    \caption{Schematic diagram illustrating components and control in a typical grid-connected single-phase PEC with GSFs system.}
    \label{fig:system}
\end{figure}

A schematic diagram of a switched single-phase voltage source PEC powered by a continuous DC bus and connected to the primary grid through an LCL filter to minimize switching noise is shown in Fig.~\ref{fig:system}. The variables $I_i$, $V_c$, $I_o$, and $V_g$ correspond to input current, capacitor voltage, output current, and grid voltage, respectively. The quadrature signal generation-based second-order generalized integral (SOGI-QSG) - phase-lock loop (PLL) is chosen due to its ease of installation and good performance~\cite{sogi2013}. The SOGI-PLL, referred to as PLL throughout the paper, is used to measure the magnitudes of the $V_g$ and $I_o$ injected by PECs while tracking the fundamental frequency ($\omega'$) and phase angle ($\theta'$).

Depending on the operational conditions, the PEC can be switched between feeding or supporting the grid depending on the operating conditions. In the former mode, the PEC injects/absorbs constant active (P) and reactive power (Q) to/from the grid, whereas, in the latter mode, it injects/absorbs variable P and Q to minimize voltage and frequency changes. The standard function and the GSFs can be used to determine the reference powers, i.e., $P^*$ and $Q^*$. The PQ controller block generates the reference current based on power references and fed into the current controller. A proportional-integral (PI) type-2 i.e., PI-2 controller, was implemented in the $d$-$q$ architecture to regulate the $I_o$ to the primary grid by the VSI. The operational power level, DC voltage, current controller parameters, and PLL parameters all influence the dynamics of this PEC system. The control method controls the grid current's dynamic responsiveness~\cite{nischal2020}. Different controllers, such as a hysteresis band controller, a Proportional-Resonant, or a PI controller, may be utilized to build the current controller~\cite{Timbus}, which will change the present response's dynamic nature. This research employs a PI-2 controller to control the current injected into the grid, where the PI gains are selected through the transient analysis using the step response~\cite{PI_controller}.

\subsection{System Identification of PECs Dynamics with GSFs}\label{SysId}

\begin{figure}[h!]
    \centering
    \includegraphics[width=\columnwidth]{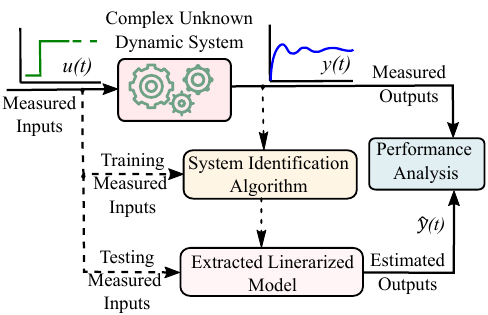}
    \caption{Basic concept of SysId. The SysId approach uses recorded input and output data to estimate unknown system dynamics using RODLTFs.}
    \label{fig:identification}
\end{figure}

SysId aims to find a continuous- or discrete-time reduced-order model that accurately reflects the system dynamics. Using SysId approaches, a mathematical model of a system that captures the dynamics of interest may be built with limited or no knowledge of the underlying control structure and parameters. Fig.~\ref{fig:identification} depicts the essential notion of a SysId process. First, the unknown dynamic system's input signal $u(t)$ and output response $y(t)$ are monitored. The training and testing datasets are then created. Using the training dataset, the SysId technique estimates model or system parameters by minimizing a given cost function, i.e., the error between the detailed system/model and the measured response. The unknown parameters are generally the coefficients of the ROFLTFs that are adjusted in the model structure such that  the norm of the predicted error is small as possible. ROFLTFs representing the detailed model/system from SysId are then validated using the testing dataset.

A series of non-linear state-space equations can be used to represent a dynamical system in a continuous-time domain as~\cite{Czop2011}:

\begin{equation}
\begin{cases}
      \Dot{x}(t) &= f(t, x(t), u(t), \omega (t);\theta)\\
      y(t) &= g(t, x(t), u(t), v(t);\theta)\\
    \end{cases}       
\end{equation}

where $f(\cdot)$ is an unknown time-varying non-linear function of the state vector $x(t)$ and control vector $y(t)$ at time $t$, and $g(\cdot)$ is an unknown time-varying non-linear measurement function; $\omega (t)$ and $v(t)$ signify the non-measurable process and measurement noise, respectively, and $\theta$ denotes a vector of unknown model parameters. The derived RODLTF model using SysId in discrete time can be written as:

\begin{equation}
\begin{cases}
      \Dot{\hat{x}}(k) &= \hat{f}(\hat{x}(k), \hat{u}(k), \hat{\omega} (k),\hat{\theta}_N(k))\\
      \hat{y}(k) &= \hat{g}( \hat{x}(k), \hat{u}(k), \hat{v}(k),\hat{\theta}_N(k))\\
    \end{cases}       
\end{equation}

\noindent where $\hat{x}(k)$ is a reduced state vector at discrete-time instance $k$, $\hat{\omega} (k)$ and $\hat{v}(k)$ signify the estimated process and measurement noise, respectively, and $\hat{\theta}_N$ denotes a vector of identified parameters using $N$ data points. The identification method utilized in this work uses the least-squares method at discrete-time instance $k$ to identify the model structure and subsequently estimate the parameters obtained by minimizing a weighted quadratic norm of the prediction errors $\epsilon(k,\theta)$, i.e., ($y(k) - \hat{y}(k|\theta))$~\cite{ljung_2012}.

\begin{equation}
    \label{eqn:cost2}
    V_N \left (\hat{y}(k),y(k),u(k), \theta \right) = \frac{1}{N} \sum_{k=1}^{N} \epsilon^2\left( k,\theta \right) 
\end{equation}
The parameter estimation of this dynamic system is determined by solving the following:
\begin{equation}
    \label{eqn:cost1}
    \hat{\theta}_N = \underset{\theta}{\text{argmin}} {\;}  V_N\left (\hat{y}(k),y(k),u(k), \theta \right)
\end{equation}



This study employs linear time-invariant (LTI) single input single output (SISO) systems. Using conversion methods such as zero-order-hold on the inputs and desired sampling time ($\text T_s$), the dynamic system model may be changed from continuous-time to discrete-time systems and vice versa~\cite{MATLABtoolbox_fit}.
Difference equations of the form in~(\ref{eqn:diffEqn}) define the relationship between the input and output on LTI discrete-time systems:
\begin{multline}
    \label{eqn:diffEqn}
    y(k) + a_1y(k-1) + a_2y(k-2) + \dots +a_ny(k-n) = \\ b_0u(k) + b_1u(k-1) + \dots +b_nu(k-n)
\end{multline}

\noindent where $u(k)$ and $y(k)$ are the system input and output, respectively, at discrete-time instance $k$. The $\mathcal{Z}$-transform of (\ref{eqn:diffEqn}) using time-shift property can be shown to be:

\begin{multline}
    Y(z) = \frac{b_0 + b_1z^{-1} + b_2z^{-2} + \dots + b_mz^{-m}}{1 + a_1z^{-1} + a_{2}z^{-2}+\dots + a_nz^{-n}}U(z) + \\ M(y_0,y_1,\dots,y_{n-1})
\end{multline}

\noindent where \textit{m} and \textit{n} are the order of the numerator and denominator, respectively, constants $[a_1~a_2\dots a_n]~\text{and}~[b_1~b_2 \dots b_m]$ are the coefficients, $M(y)$ is an initial condition term. The rational expression multiplying discrete-time input signal $U(z)$ is the discrete input of a RODLTF.

\subsection{Probing Signals}\label{Prob_sig}

Perturbing PECs with correctly designed probing signals to obtain information about system behavior is vital in data-driven modeling. The system may have different time constants, so the probing signal should be designed to capture a wide range of system dynamics across multiple operating modes. The accuracy and generalization of the model depends on how the active probing signals are designed. Several significant design considerations of probing signals can be found in detail for classical techniques in~\cite{pierre2010signals}. The input signal amplitude should cover the entire range from minimum ($V_{low}$) to maximum ($V_{max}$) to optimize the input signal's strength and, as a result, the signal-to-noise ratio. In addition, the power content of the input signal should focus on the frequency spectrum narrowed to the desired dynamic range that the model will be used, thus generating a greater model accuracy within this range.

This study uses a broadband chirp signal to perturb the PV PEC system and extract the RODLTF model to achieve the aforementioned objectives. Logarithmic, square chirp signals have outperformed other signals (square, sine, logarithmic sine chirp) in extracting the PEC dynamics to fit a data-driven RODLTF model in~\cite{nischal2022speedam_pv_inv_modeling}. The selected logarithmic, square chirp signal is a logarithmic frequency sweep of a square wave signal and is generated as in Eqs. (\ref{eq:chirp1}) and (\ref{eq:chirp2}) utilizing SysId theory (i.e., the content of the probing signal in the frequency band) and the power system's design constraints, which are primarily determined by three factors: energy, amplitude, and period.

\RestyleAlgo{ruled}
\begin{algorithm}[hbt!]
\caption{Probing Signal Generation}\label{alg:probing_signal}
\textbf{Input:} Operational limits $\gets [V_{low}, V_{max}]$\
\text{Number of partitions $\gets \mathcal{N} \in [\mathcal{N}_{min} ~ \mathcal{N}_{max}]$}\\

 \textit{Initialization} : 

Time interval for each partition $\gets T$\

Simulation clock time $\gets t$\

Simulation run time $T_{run} \gets T \times \mathcal{N}$\

\While{$T_{run} \neq t$}{
    $k \gets floor(t/T)$\
    
    \vspace{3 pt}
    $dV \gets \frac{V_{max}-V_{low}}{\mathcal{N}}$\
    \vspace{3 pt}
    
    $V_1 \gets V_{low}  + k \times dV$\
    
    $V_2 \gets V_1  + dV$\
    
    {\If{$V_2 > V_{max}$}{
      $V_1 \gets V_{max}  - dV$\
      
      $V_2 \gets V_{max}$\
    }
  }
}
\end{algorithm}

The signal, along with its algorithm that was used in the research, has the following fundamental equation:
 \begin{equation}
 \label{eq:chirp1}
     x(t) =  A \text{cos}(2\pi f(t) + \phi_0)
 \end{equation}
\noindent where $\phi_0$ is the starting phase angle, $A$ is the amplitude, and $f(t)$ is the frequency of the signal at time $t$. The frequency sweep $f(t)$ is defined as

 \begin{equation}
 \label{eq:chirp2}
     f(t) =  f_0 \times \frac{f_0}{f_1} ^ {{(\frac{1}{T}})^{t}}
 \end{equation}
\noindent where $f_0$ denotes the starting frequency, $f_1$ denotes the end frequency, and $T$ is the chirp signal sweeping time. With the selected number of input partitions $\mathcal{N}$ and operating limits $[V_{low} ~ V_{max}]$, the desired signal is generated based on Algorithm~\ref{alg:probing_signal}.

\subsection{Metrics for quantitative assessment}\label{Metrics}

After the data is collected and the model is fit, it is required to validate the discovered model by evaluating its statistical performance. A collection of models with varying numbers of poles and zeros can fit the data based on the acquired input-output data. A goodness-of-fit ($\textit{fitpercent}$) measure based on the NRMSE is used to calculate the model's fit percentage as~\cite{MATLABtoolbox_fit}:

\begin{equation}
    \label{eqn:fitpercentage}
    \textit{fitpercent} = 100\times\left(1 - \text{NRMSE}\right)\%
\end{equation}

\noindent where $||.||_2$ in NRMSE indicates the 2-norm of the vector, $y(t)$ denotes data from the detailed PEC simulation model, $\hat{y}(t)$ denotes data from the aggregated RODLTFs simulation model, and $\bar{y}(t)$ indicates the channel-wise mean of the measured output data. The percentage fit in~(\ref{eqn:fitpercentage}) indicates how well the response of the derived models fits the estimation data. The quality of the models with acceptable poles and zeroes is evaluated using additional metrics adjusted $R^2$, AIC, and BIC, which trade-off the goodness of fit and model complexity. The most accurate model with the highest value of corrected goodness-of-fit (i.e., Adj. $R^2$) and the lowest value of AIC and BIC is selected as the final RODLTF model. The details of these metrics are summarized in the Appendix in Table~I.

\section{Proposed Partitioned Modeling Approach}\label{partitioned_model}

Existing converters are being complemented with new IEEE 1547-2018 standard-based GSFs such as voltage-VAr/Watt, frequency-Watt, ramp-rate control, and voltage/frequency ride through to support power system voltage and frequency. The dynamics of PECs with diverse modes of operation change throughout the power system, prompting the development of novel modeling methodologies for proper system design, operation, and dispatch. This study presents partition model development and the speedup of the proposed approach  under the Volt-VAr mode of operation based on IEEE 1547-2018; however, this approach applies to other modes and standards.

\subsection{Statistical Partitioned Model Development}
\begin{figure}[h!]
    \centering
    \includegraphics[width=\columnwidth]{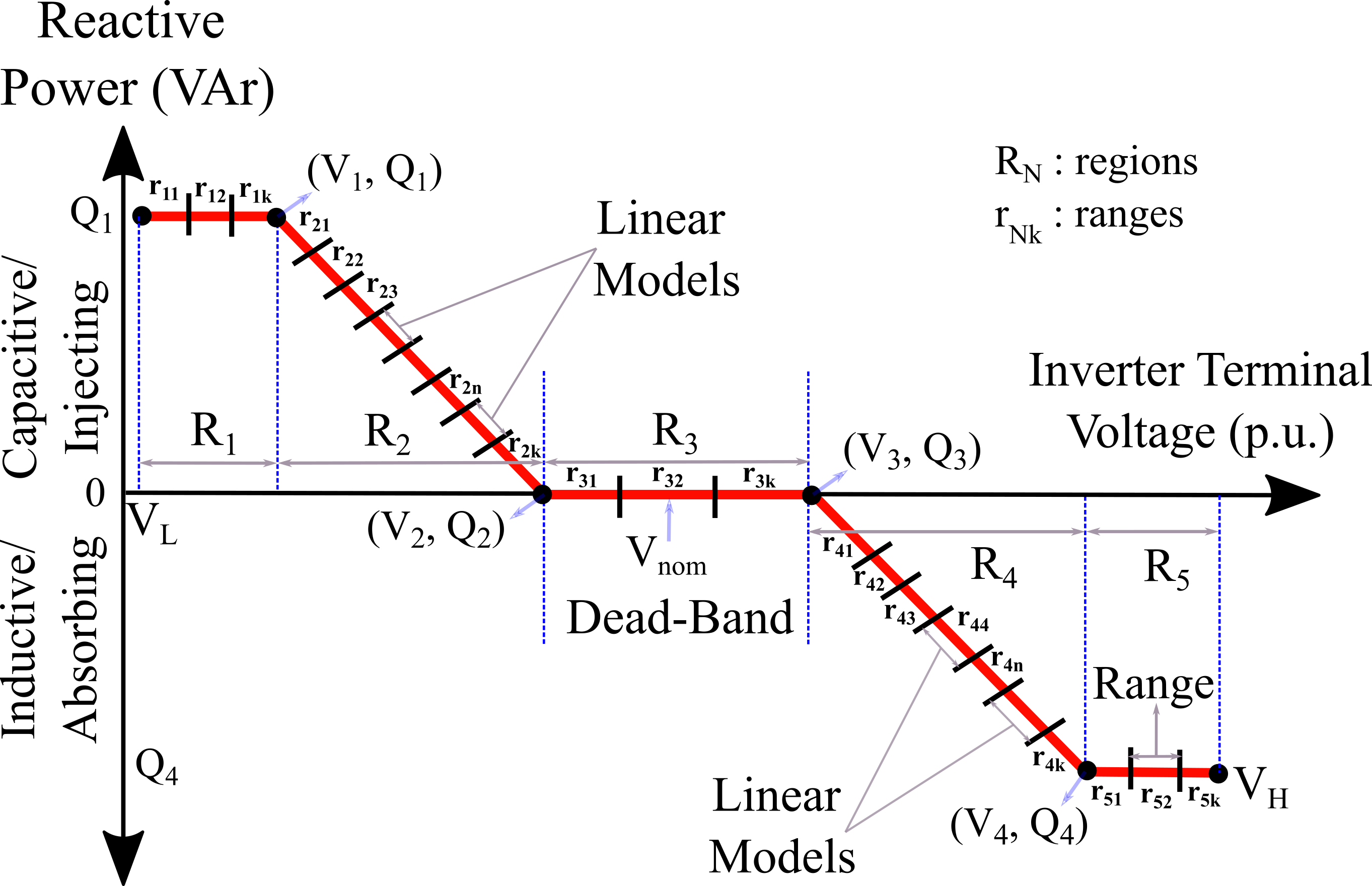}
    \vspace{2 pt}
    \caption{The graphic demonstrates the Volt-VAr piecewise linear characteristics curve.}
    \label{fig:volt_var}
\end{figure}

The IEEE 1547 Volt-VAr mode adjusts the PEC's reactive power output in response to grid voltage changes. In this mode, the new IEEE standard 1547-2018 mandates DERs to actively modify reactive power output as a function of the voltage using a piecewise linear characteristics curve. The Volt-VAr curve is shown in Fig.~\ref{fig:volt_var}, where the form of the curve is determined by the change in reactive power parameters $Q_1$, $Q_2$, $Q_3$, and $Q_4$ with voltage parameters $V_1$, $V_2$, $V_3$, and $V_4$. The equation for the piecewise characteristic is expressed as:

\begin{align}
\label{volt-var}
    Q=\begin{cases}
    Q_1 & \text{if }V_L \leq V < V_1\\
    Q_1+\frac{Q_2-Q_1}{V_2-V_1}(V-V_1) &\text{if } V_1 \leq V < V_2\\
    0 &\text{if } V_2 \leq V < V_3\\
    Q_3 + \frac{Q_4-Q_3}{V_4-V_3}(V-V_3) &\text{if } V_3 \leq V < V_4\\
    Q_4 &\text{if }V_4 \leq V < V_H
    \end{cases}
\end{align} 

where $V_H$ and $V_L$ are the upper and lower limits for DER continuous operation. When $V>V_H$ or $V<V_L$, the voltage ride-through feature is enabled after this function has disengaged. As long as the reactive power does not exceed the PEC limit, the reactive reference power ($Q_{ref}$) can be written as:

\begin{align}
\label{eqn4}
    Q_{ref} = \text{min}\left(Q, \sqrt{S^2-P_{inv}^2}\right)
\end{align}

Referring to Fig.~\ref{fig:volt_var}, when the PEC terminal voltage is between $V_2$ and $V_3$, the reactive power output of the PEC is zero, and the region inside that voltage range is known as the ``dead-band''. If the grid voltage falls below the lower dead-band limit of $V_2$, reactive power is injected into the grid until the grid voltage falls below $V_1$. When the grid voltage reaches the upper dead-band limit of $V_3$, reactive power is absorbed; constant reactive power is absorbed when the grid voltage surpasses $V_4$.


The dynamics of a converter's operating regions (within each GSF) can be described using multiple dynamic, computationally efficient linear models. To achieve both speed and accuracy, we propose a partitioned linearized modeling approach in which discrete regions of modes of operation (e.g., Volt-VAr/Watt) are subdivided into smaller ranges. In contrast to our earlier work~\cite{sunil2021}, where we achieved the recommended technique using evenly split voltage magnitudes, we have updated the algorithm using a binary search and selected the appropriate number of partitions using the statistical metrics from Table~I. The detailed steps are given in Algorithm~\ref{alg:partitioning_modeling}.

\RestyleAlgo{ruled}
\begin{algorithm}[hbt!]
\caption{Partitioned Modeling with Binary Search}\label{alg:partitioning_modeling}
\textbf{Input: Minimum \& maximum partitions.} in\\
\textbf{Output: Optimal number of partitions.} out\\
 \textit{Initialization} : number of partitions = $[\mathcal{N}_{min} ~ \mathcal{N}_{max}]$, voltage limits = $[V_{min} ~ V_{max}]$, \textit{fitpercent} = $fit_{req}$\\

Calculate overall \textit{fitpercent} for $\mathcal{N}_{min}$ and $\mathcal{N}_{max}$
\\
check = True \\
\While{$\mathcal{N}_{max} - \mathcal{N}_{min} > 1$}{

    {\eIf{$fit_{\mathcal{N}_{max}} > fit_{\mathcal{N}_{min}}$ and $fit_{\mathcal{N}_{max}} > fit_{{req}}$}{
      $\mathcal{N}_{mid} \gets \mathcal{N}_{min} + (\mathcal{N}_{max}-\mathcal{N}_{min})//2$\\
      \vspace{1 em}
      \begin{enumerate}
          \item Run MATLAB API with $\mathcal{N} = \mathcal{N}_{mid}$\\
      \item Run System Identification Algorithm\\
      in Python for $fit_{\mathcal{N}_{mid}}$\\
      \end{enumerate}

      \vspace{1 em}
      
         {\eIf{$fit_{\mathcal{N}_{mid}} >= fit_{req}$}{
      $\mathcal{N}_{max} \gets N_{mid}$\\
      $fit_{\mathcal{N}_{max}} \gets fit_{\mathcal{N}_{mid}}$}{
      $\mathcal{N}_{min} \gets \mathcal{N}_{mid}$\\
      $fit_{\mathcal{N}_{min}} \gets fit_{\mathcal{N}_{mid}}$\
    }
  }
    }{$fit_{\mathcal{N}_{min}} > fit_{\mathcal{N}_{max}}$\\
    check = False\\
    break}
  }

}
 ${\mathcal{N}} = {\mathcal{N}_{max}}$ \textbf{if} check \textbf{else} ${\mathcal{N}_{min}}$\\
 $fit_{\mathcal{N}} = {fit_{\mathcal{N}_{max}}}$ \textbf{if} check \textbf{else} ${fit_{\mathcal{N}_{min}}}$
\end{algorithm}

Two instances of the MATLAB API are started for $\mathcal{N}_{min}$ and $\mathcal{N}_{max}$. A signal is generated for each instance, and data is collected. To obtain the RODLTF models for each minimum and maximum partition using the SysId approach, two Python instances are utilized. Performance is measured using the defined error metrics. After acquiring the $\textit{fitpercent}$ for both instances, we compare and choose whether or not to activate the partitions. Once the condition is met, the MATLAB API generates signals and retrieves data for $\mathcal{N}_{mid}$, and Python applies the SysId, respectively. This process is repeated until the stopping condition is not met.

\section{Case Study in a Single-House System}\label{casestudy_singlehouse}
\subsection{Co-Simulation Setup}

\begin{figure*}[!h]
    \centering
    \includegraphics[width=\textwidth]{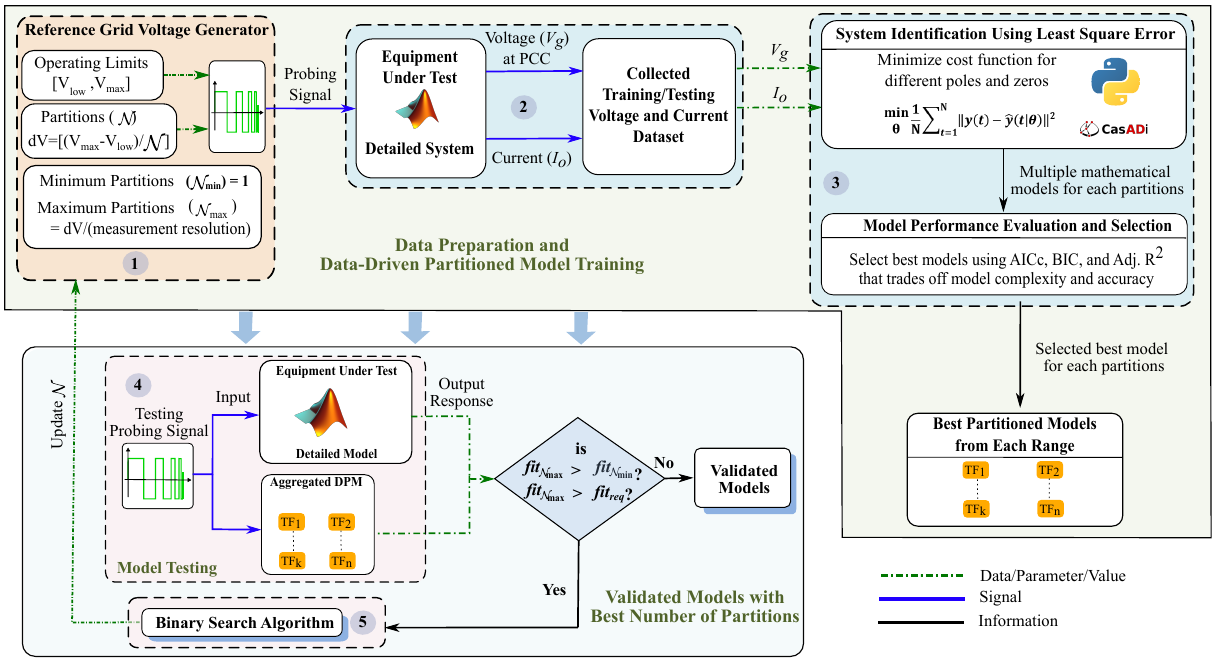}
    \caption{Schematic flow diagram of data-driven partitioned model training, testing, and validation using a co-simulation framework.}
    \label{fig:FLOW_diag}
\end{figure*}

This section describes the co-simulation setup for the development of partitioned models of a PEC. Fig.~\ref{fig:FLOW_diag} shows the overall flow to identify PEC's linear partitioned RODLTF models with GSFs. To determine the dynamic linear models, a logarithmic, square chirp signal of varying voltage amplitude was used to perturb the PEC. Though this approach is applicable to extract the dynamics of any PEC having any GSFs, this study shows the extraction of PEC dynamics in Volt-VAr mode. In this study normal operating range, i.e., 0.88 to 1.1 p.u., is taken as $V_{min}$ and $V_{max}$, respectively. The minimum number of partitions was set to $\mathcal{N}_{min}=1$, and the maximum number of partitions was set to $\mathcal{N}_{max}=22$. This number was chosen because the measurement noise dominates the input after a measurement resolution of 0.01. 


A binary search algorithm was used to determine  $\mathcal{N}$, resulting in the best \textit{fitpercent}. First, the signal was generated for $\mathcal{N}_{min}$ and $\mathcal{N}_{max}$ using Algorithm~\ref{alg:probing_signal} and based on the input-output measurements of a PEC operating in Volt-VAr mode, which were gathered in response to a probing signal, the SysId algorithm was applied to identify the linear models that minimized the cost function for different numbers of model poles and zeros. Then the models are evaluated with the goodness-of-fit and are selected based on the information criterion illustrated in Table~I. The cost function in~(\ref{eqn:cost1}) is solved using the ``IPOPT" solver, a non-linear programming open-source primal-dual interior point method interfaced with CasADi~\cite{casadi}.

\begin{figure}[ht!]
    \centering
    \includegraphics[width=\columnwidth]{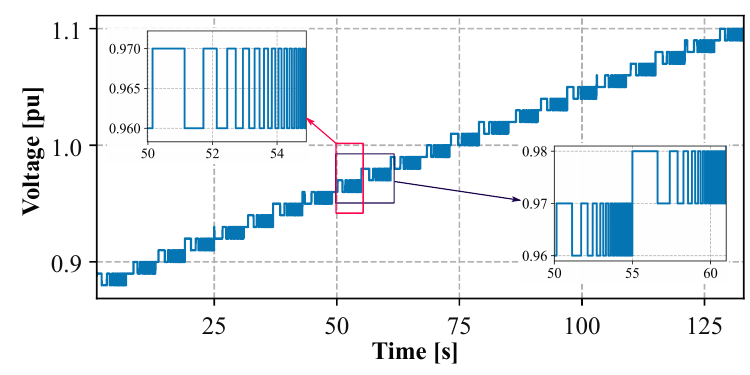}
    \caption{Snapshot of logarithmic, square chirp signal as the probing signal used to perturb the PEC with Volt-VAr support.}
    \label{fig:sweep1}
\end{figure}

The snapshot of the probing signal used to perturb the PEC is shown in Fig.~\ref{fig:sweep1}. The PEC is perturbed within the normal operating IEEE voltage limits, i.e., 0.88--1.1 p.u. The voltage at the PCC was perturbed by using a logarithmic, square chirp signal (Fig. \ref{fig:sweep1}) with voltage amplitude varied from 0.88 p.u. to 0.89 p.u. for 6 s, and then the amplitude was increased by 0.01 p.u. for the following range until it reaches 1.1 p.u. The step-change of 0.01 p.u. is typically used by relying on the availability of the voltage within the specified range (0.88 p.u.-1.1 p.u.). The frequency is swept between 1 and 32 Hz and was chosen based on the PEC's settling time response.

\begin{figure*}[ht!]
    \centering
    \includegraphics[width=0.95\textwidth]{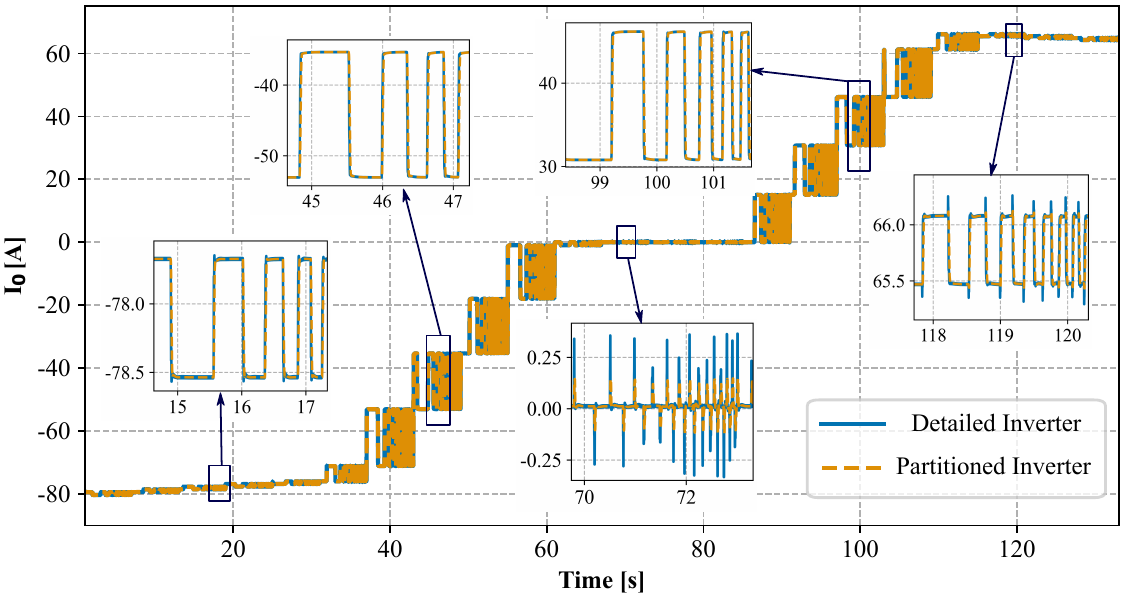}
    \caption{Response of the output current from detailed PEC model compared to the output current from partitioned PEC model.}
    \label{fig:validation}
\end{figure*}

\subsection{Model Validation}
The implementation of the partitioned modeling with a binary search algorithm, along with its performance in a modified radial distribution system, will be analyzed in this section. First, the Volt-VAr characteristics curve was activated according to the Volt-VAr setting specified in Table ~II.

The voltage at PCC was perturbed using the square chirp signal. After perturbation, the actual response of the inverter is noted as the response of the detailed inverter as depicted in  Fig.~\ref{fig:validation}. It can be noted that $I_0 [A]$ is increasing (which means reactive power is decreasing ($I_q$ and $Q$ are 90\degree~apart, i.e., these two signals are opposite during representation, meaning when Q is declining, then $I_o$ [A] seems increasing and vice versa). As per Fig.~\ref{fig:sweep1}, the voltage is rising, which indicates that reactive power should be decreased ($I_0$ will be increased) to support the voltage during Volt-VAr mode, which happens as assumed and is shown as the detailed inverter in Fig.~\ref{fig:validation}. It can be noted that $I_0$ increases from the initial time to 35~s with a slight slope. However, at 35 s, the Volt-VAr mode of the inverter is activated (voltage exceeds 0.92 p.u.), and the inverter reacts accordingly to the Volt-VAr curve with a defined gradient until it reaches 0.98 p.u. at 65~s. After exceeding 65~s to 85~s, the inverter reaches the dead band between 0.98 p.u. - 1.02 p.u., and reactive power supplied by the inverter is near-zero. After 85~s, the inverter absorbs reactive power until 110 s, where the voltage amplitude reaches 1.08 p.u. and the absorption reaches a constant maximum. After 110 s, the Volt-VAr mode is deactivated as the voltage exceeds the allowable 1.08 p.u., and the inverter does not absorb/consume any reactive power. However, due to the Volt-VAr mode saturation, the PEC tries to limit the voltage  with constant reactive power beyond 1.08 p.u. and below 0.88 p.u.

\begin{table}[!h]
    \centering
    \caption*{Table II \\ Volt-VAr Parameters.}
    \label{tab:inverter_parameter}
\begin{tabular}{cc}
\hline
\multicolumn{1}{c}{\textbf{Parameter}}      & \textbf{Value}                        \\ \hline                        
\multicolumn{2}{c}{\textbf{Volt-VAr Setting}}                                               \\ \hline\hline
\multicolumn{1}{c}{$Q_1$, $Q_2$, $Q_3$ and $Q_4$} &  6.25, 0, 0, and -6.25 kW                 \\ \hline
\multicolumn{1}{l}{$V_L$, $V_1$, $V_2$, $V_3$, $V_4$, and $V_H$ } & 0.88, 0.92, 0.98, 1.02, 1.08, 1.10 p.u. \\ \hline
\end{tabular}
\end{table}

The SysId algorithm developed in Python is employed, which utilizes 70\% of Voltage (p.u.) and its corresponding response $I_{0}$ (A) as input and output data, respectively, for obtaining the RODLTF of inverter dynamics. The partitioned modeling with a binary search algorithm is then implemented in Python, which generates a large number of RODLTF by varying the number of poles and zeros and calculates the \textit{fitpercent} of RODLTF. Then it generates the best RODLTF for each range based on \textit{fitpercent}. The selected models also correspond to the models with the highest Adj. R$^2$ and the lowest AICc and BIC. The remaining 30\% of Voltage (p.u.) as input is then passed through the obtained RODLTF, and the output is determined for cross-validating. The partitioned inverter model of Fig. \ref{fig:validation} is the response of the inverter after applying partitioned modeling and cross-validating the obtained RODLTF. 

\begin{figure*}[ht!]
    \centering
    \includegraphics[width=0.95\textwidth]{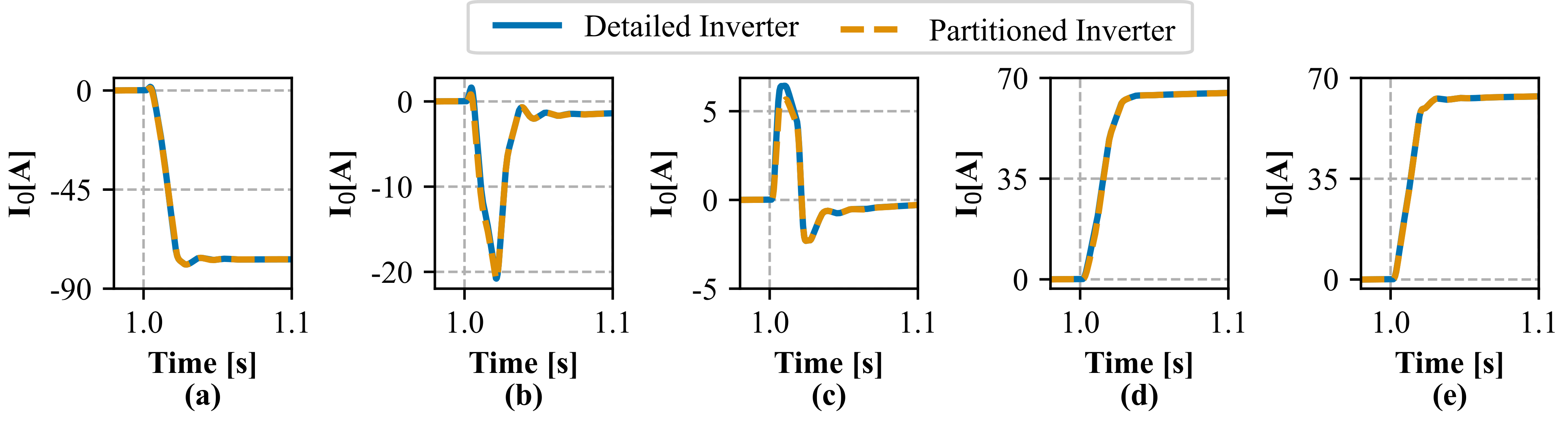}
    \caption{Response from detailed and partitioned inverter model with step signal as input in each PEC Volt/VAr operating region. (a) Response with step signal from 0.88 to 0.92 p.u. (b) Response with step signal from 0.92 to 0.98 p.u. (c) Response with step signal from 0.98 to 1.02 p.u. (d) Response with step signal from 1.02 to 1.08 p.u. (e) Response with step signal from 1.08 to 1.1 p.u.}
    \label{fig:Step_Response}
\end{figure*}

To further validate the derived partitioned models of smart inverter operating under Volt-VAr mode, a unit step response is given to both the detailed and partitioned inverter models, and the response in the form of output current is compared as shown in Fig.~\ref{fig:Step_Response}. The step input to the system is given in such a way that it covers the IEEE-defined Volt-VAr curve set-points while ensuring that all the derived partitioned inverter models activate. The step input with the first set-point range of 0.88 to 0.92 p.u is used for both the detailed and the partitioned inverter model, and the output inverter current is compared. Similarly, step input is changed from 0.92 to 0.98 p.u., 0.98 to 1.02 p.u. , 1.02 to 1.08 p.u., and 1.08 to 1.1 p.u., and the output response is recorded. The results show that the transient dynamics from the derived partitioned inverter model closely follow the detailed inverter transient dynamics.

\section{Case Study in a Modified Radial Distribution System}\label{casestudy_twelvehouse}

A study was carried out in this section to assess the developed reduce-ordered linear dynamic smart inverter models operating in Volt-VAr mode in multiple residential homes with varied load scenarios in accordance with the benchmark's specifications. Although this proposed framework applies to any system, to test the scalability of our partitioned PEC model, this study uses a 12-house radial distribution benchmark system from~\cite{Reinaldo2011} implemented in MATLAB/Simulink. 

\subsection{Modified Distribution Benchmark Model}

\begin{figure}[ht!]
    \centering
    \includegraphics{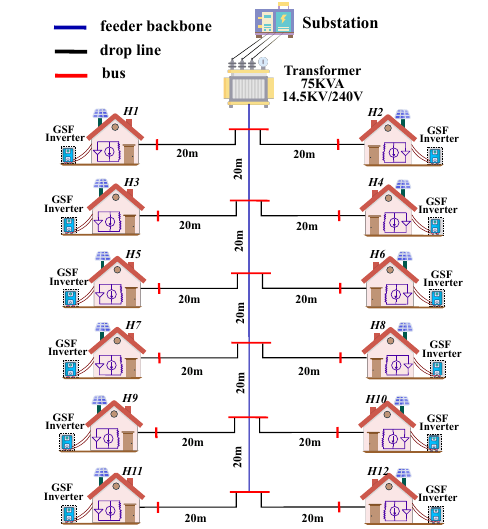}
    \vspace{1 em}
    \caption{Modified overhead 12-residential house test feeder with ZIP load and 8.4 kW grid-connected GSF PV PEC.}
    \label{fig:benchmark}
\end{figure}

 Each house from the benchmark systems is modified to have dynamic ZIP loads and assumed to have a grid-connected PV PEC with GSFs from IEEE 1547-2018 standard with an installed peak PV capacity of 8.4 kW. As depicted in Fig.~\ref{fig:benchmark}, the residences are supplied via a 75 kVA single-phase 14.4 kV–120/240 V distribution transformer with the split-phase supply of 1.02 p.u. at the secondary side to allow a maximum 5\% voltage drop at the last customer ($H_{11/12}$). The total feeder length is 120 m, the feeder backbone line is 100 m long, and each house is connected with a 20 m drop line. Table~III and Table~IV in the Appendix show the transformer and feeder parameters, respectively taken from~\cite{Reinaldo2011}. The voltage limits established for DER protection are 0.88 to 1.1 p.u. and reconnection thresholds are 0.89 to 1.09 p.u.

\subsection{Load Models}
As a result of advances in power electronics, significant changes in appliance load characteristics have occurred in recent years. Experiments in the laboratory on a wide range of appliances and equipment have shown no load is truly constant-impedance (Z), constant-current (I), or constant-power (P)~\cite{cvr_zip_2013}. The ZIP coefficients have been used to express the P-V and Q-V characteristics of electric loads in the system to characterize the load dynamics~\cite{zipload1995}. The load composition of customers in a given service class influences the P-V and Q-V curves. The parameter of ZIP load models is calculated by fitting the test data from voltage reduction laboratory tests with a least-square fitting algorithm~\cite{bokahari_zip2014}. ZIP coefficients model can be written with the following quadratic expressions:

\begin{equation}
\label{eqn:zip_p}
   P = P_{0} \left[ Z_{p} \left( \frac{V}{{V_{0}}} \right)^{2} + I_{p} \frac{V}{{V_{0}}} + P_{p} \right]
\end{equation}

\begin{equation}
\label{eqn:zip_q}
  Q = Q_{0} \left[ Z_{q} \left( \frac{V}{{V_{0}}} \right)^{2} + I_{q} \frac{V}{{V_{0}}} + P_{q} \right]
\end{equation}

Active power dynamics ($P-V$) are represented by~(\ref{eqn:zip_p}) with $Z_p$ the constant impedance coefficient, $I_p$ the constant current coefficient, and $P_p$ the constant power coefficient.  The reactive power dynamics ($Q-V$) are obtained by~(\ref{eqn:zip_q}) using the parameters $Z_q$, $I_q$, and $P_q$. In (\ref{eqn:zip_p}) and (\ref{eqn:zip_q}), $V$, $V_0$, and $P_0$ are the local voltage, nominal voltage, and reference $P$ at $V_0$. Similarly $Q_0$ is the reference $Q$ at $V_0$. The ZIP coefficients must meet the following two conditions:

\begin{equation}
    Z_p + I_p + P_p = 1 ~\&~ Z_q + I_q + P_q = 1
\end{equation}

\subsection{Solar Irradiance and Load Profile}
\begin{figure}[h!]
    \centering
    \includegraphics[width=0.9\columnwidth]{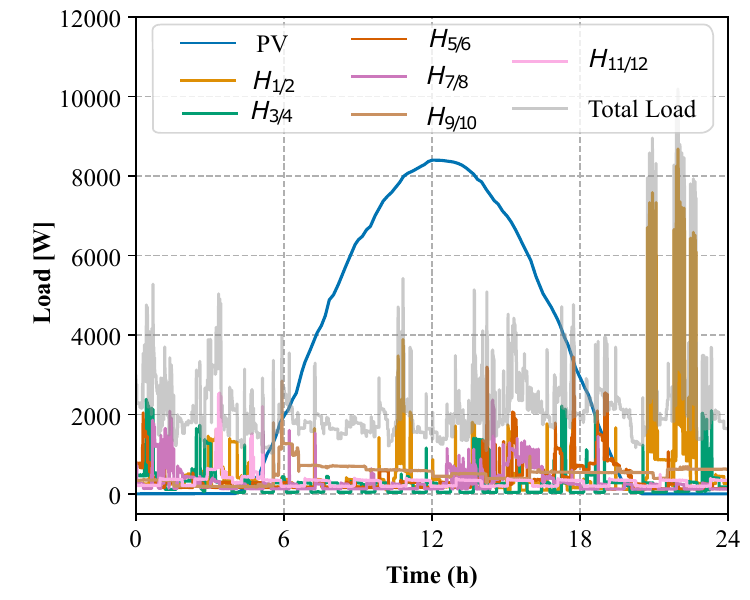}
    \caption{Voltage profile in low voltage modified feeder of each house.}
    \label{fig:load_PV}
\end{figure}

The load data was obtained from Residential Energy Disaggregation Dataset (REDD)~\cite{REDD}, a publicly available data set that contains the power consumption from real homes of California regions. The resolution of one second was taken for the daily household load profiles. Based on the individual appliance's power consumption and load profile for the residential consumer class, ZIP coefficients for stratum D out of six stratum were taken from~\cite{bokahari_zip2014}, and are shown in the Appendix in Table~IV. 

The solar irradiance data from Chicago, IL, 2014 was taken from SolarAnywhere~\cite{solaranywhere_2022}. The solar data required for one-second resolution was calculated using linear interpolation to form the hourly data. The irradiance data was used to calculate the maximum available power in the PV array as $P_{MPPT} = \mu \times I \times A$ where efficiency ($\mu$) and area  (A) of PV panel was 16.7 $\%$ and \text{50.2605 ($m^2$)} respectively, and I is irradiance in \textbf{$W/m^2$}. The total residential load, solar power data, and individual residential load profiles for a day are presented in Fig.~\ref{fig:load_PV} to generalize the diverse load profiles of the twelve households.

\subsection{Partitioned Model Speedup}

Due to the increased computational complexity in representing detailed dynamic models of numerous IBRs interacting with the power system, it is challenging to model the dynamics of CDPS due to the long simulation time. One approach we utilized is to compile an optimized  C-version of the Simulink model. The model can be run with C-interfaced Python to speed up the simulation process further. To interface the Simulink model to Python, the following steps were executed:
\begin{enumerate}
  \item Generate C-code using the C-code generation tool from Simulink. 
  \item Generate a new interface C-code, header file, and Python code to interface the previously generated C code from Simulink to Python. 
  \item Compile to a shared C library that includes and links all the main functions, sources, and header files. 
  \item Compile using the GCC C compiler.  
  \item Import the interface code and perform simulation in Python.
\end{enumerate}

\section{Results Analysis}\label{Results_and_analysis}
The modified benchmark model is initially validated against the benchmark provided in~\cite{Reinaldo2011} by sweeping the system loading. Once validated, a detailed analysis of the Volt-VAr mode activation results in the modified 12-house system is conducted, and the performance of the partitioned modeling technique-produced reduce-ordered linear dynamic models under various loading and generating situations was compared. The simulation was performed on an Intel(R) Core(TM) i7-2600 CPU @ 3.40GHz computer with 16 GB RAM.

\subsection{Performance With Load Sweep and Constant Irradiance}

\begin{figure}[h!]
    \centering
    \includegraphics[width=0.9\columnwidth]{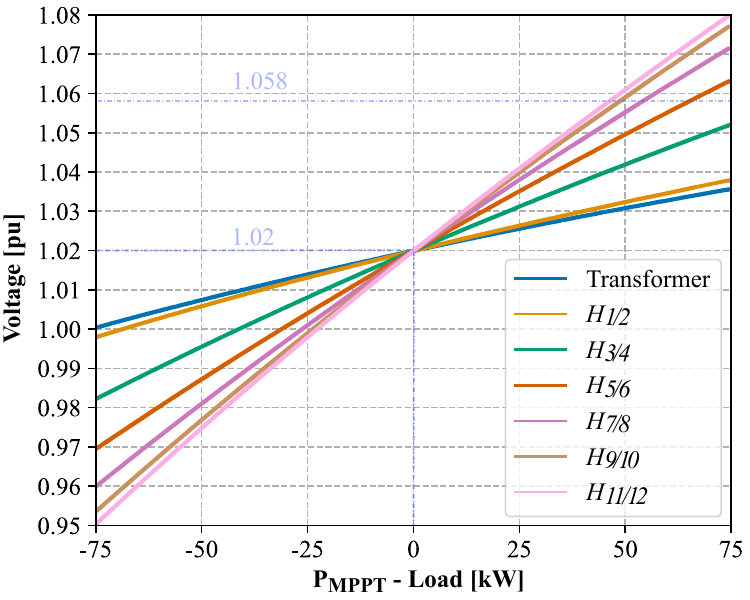}
    \caption{Voltage profile in low voltage modified feeder of each house.}
    \label{fig:sweep}
\end{figure}

In the base scenario, the loading/generation of the modified system is assumed to vary between $\pm$75 kW ($\approx$ 6.25~ kW$\times$12) at unity power factor. The model in Fig.~\ref{fig:benchmark} was simulated without activating the GSFs, where the PV PEC injects maximum power until the terminal voltage $V_H$ (i.e., 1.10 p.u.) is reached. Fig.~\ref{fig:sweep} shows the net load sweep without any control results in the same voltage profile for houses at the same distance. Moreover, if net loading/generation is zero, the secondary side is 1.02 p.u. From the results, only $H_{1}$ to $H_{4}$ undergo typical operating voltages as defined by American National Standard Institute (ANSI), i.e., 1.058 p.u. The remaining houses experience overvoltage during the 24-hour simulation period similar to~\cite{Reinaldo2011}. Given the unpredictability of the system's loads, this ANSI limit can be interpreted as the maximum quantity of PV that can be put in the feeder without over-voltage protection devices.

\subsection{Performance With Different House Loads}

\begin{figure}[h!]
    \centering
    \includegraphics[width=0.9\columnwidth]{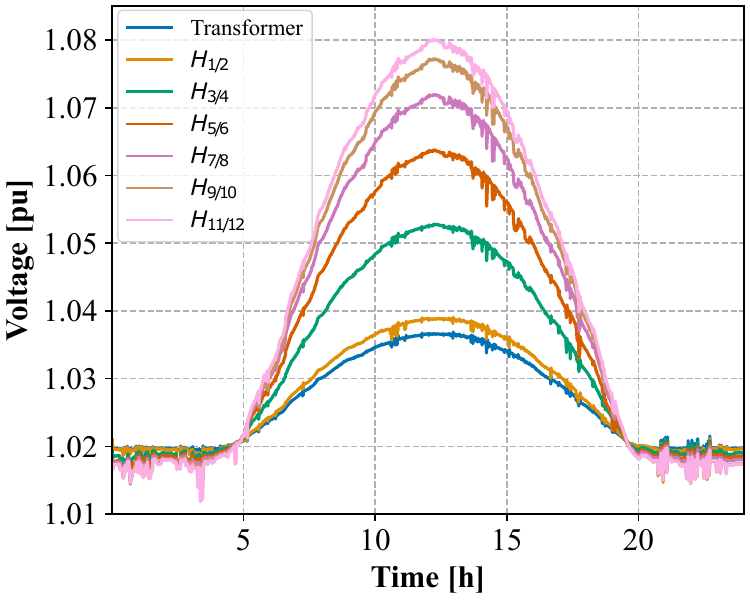}
    \caption{A 24-h voltage profile without Volt-VAr support for over-voltage prevention.}
    \label{fig:wgsf}
\end{figure}

The typical PV generation and the actual load profiles must be considered, as well as the voltage profile and the amount of reactive power curtailed using the PV PEC with GSFs for the LV distribution feeder. This section applies to a period of 24 hours. Fig.~\ref{fig:wgsf} depicts the voltage profile in the base scenario without Volt-VAr activation. The results show the voltage profile in twelve residences and the secondary side of the transformer. Because of the higher resistance in the line from the transformer, the voltage in houses connected to the feeder at a longer distance is higher for the same current injection. 

The system's maximum voltage of (1.08 p.u.) is reached at midday in $H_{11/12}$ at the end of the feeder; $H_{5}$ to $H_{12}$ experience over-voltage ($>$ 1.058 p.u.) for at least a short period during the day. Based on these results, because the average loading in the system is very low ($\approx$ 4~kW while the total generation is higher, the PV capacity should be limited to about $4.5$~kW to prevent over-voltage if no GSFs are implemented.

\begin{figure}[!h]
    \centering
    \includegraphics[width=0.9\columnwidth]{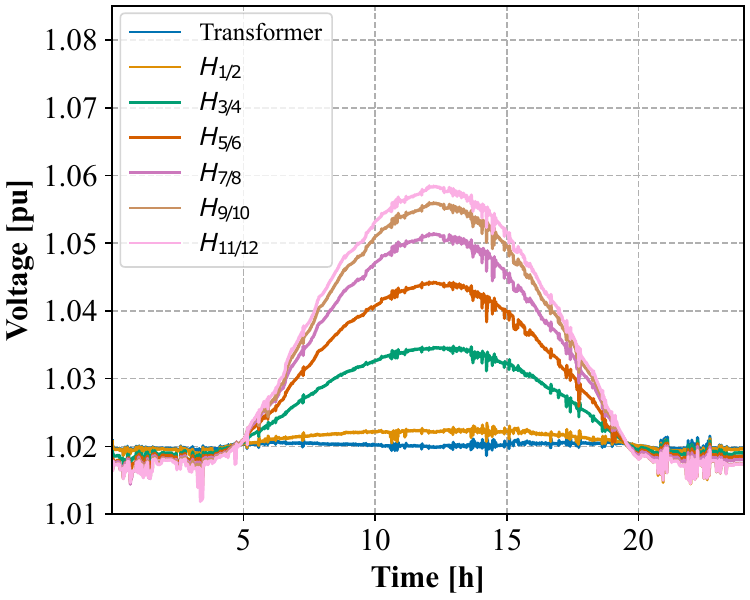}
    \caption{A 24-h voltage profile with Volt-VAr support for over-voltage prevention.}
    \label{fig:gsf}
\end{figure}

Fig.~\ref{fig:gsf} presents the voltage profile with the PEC using Volt-VAr. Houses $H_{11}$ and $H_{12}$ experienced a voltage just above 1.058 p.u. The maximum voltage in the system (1.059 p.u.) occurs at noon, in $H_{11/12}$, after the voltage has been reduced by almost 0.021 p.u. with the activation of the Volt-VAr mode. However, the last houses still experience over-voltage due to the higher net generation. Moreover, the voltage in $H_{1}$ to $H_{10}$ is reduced below the ANSI limits (1.058 p.u.). Even with the absorption of reactive power, the voltage at the secondary side of the transformer maintained around 1.02 p.u. The over-voltages could be further prevented by turning on Volt-Watt, which is left for a future exercise.



\begin{figure}[h!]
    \centering
    \includegraphics[width=0.9\columnwidth]{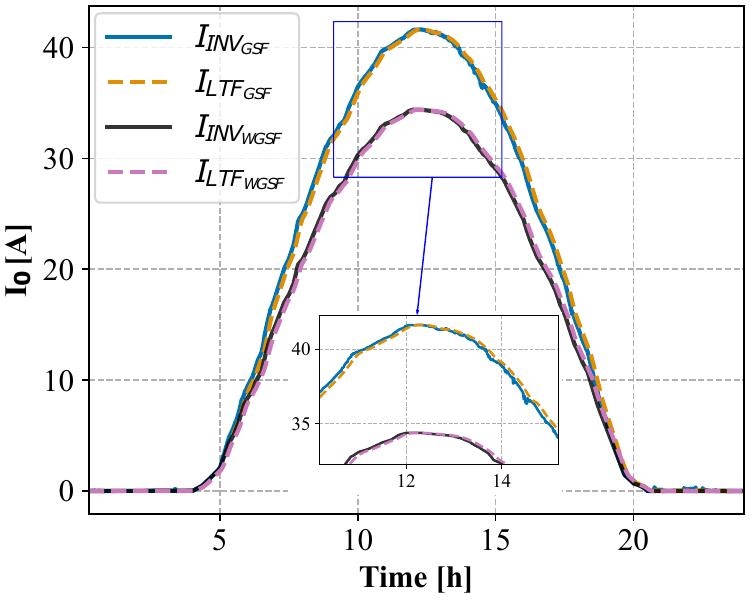}
    \caption{Comparison of current injected by derived RODLTF model and the detailed PEC model with Volt-VAr (GSF) and without Volt-VAr (WGSF) activation in $H_{11/12}$.}
    \label{fig:curr}
\end{figure}

Finally, for performance analysis, the partitioned RODLTF model is compared with the detailed PEC model, and the current injected by the PEC is used for the comparison. Fig.~\ref{fig:curr} represents the current injected by the partitioned model and the detailed PEC model in the farthest houses, i.e., $H_{11/12}$ from the transformer. The result shows that without the Volt-VAr mode activation, the current injected by the partitioned PEC model closely follows the current injected by the detailed PEC model with an average NRMSE of 0.67$\%$. In contrast, in the case of activating Volt-VAr mode, the average NRMSE is found to be 1.98$\%$.

In addition to comparing the outcomes of the two simulation models, the simulation time was evaluated. The partitioned linearized PEC model resulted in a 2.5$\times$ speedup compared to  the detailed PEC model. The Simulink model was further compiled into C-code, and a Python interface was created to speed up the simulation further. Using this approach, the derived model was 6.5$\times$ faster than the detailed model. As the number of PEC inverters increases, this framework provides a promising technique to run more extensive simulations for voltage dynamics at a lower computational cost with acceptable accuracy.

\section{Conclusions}\label{conclusion}
This paper presents the automated data-driven extraction and validation of PV PEC dynamics with the GSFs. The proposed partitioned modeling algorithm was shown to derive the RODLTF of PEC operating at Volt-VAr mode. A single-house PV PEC model with GSF was first used to test and validate the derived partitioned model. The modified benchmark from a Canadian suburban distribution feeder was used to demonstrate scalability. It was shown that the PEC model derived using the proposed partitioned algorithm could accurately represent the dynamics of the PEC operating at Volt-VAr mode with an accuracy of 97\%. Moreover, the derived model was used in the modified benchmark model, and the results show that these models can be expanded to study more extensive systems with faster (i.e., 6.5 $\times$) simulation time compared to a detailed PEC model. While conducting the comprehensive CDPS dynamic studies, the use of the faster approach that has been proposed could lower the computational expense.

\section*{Acknowledgement}
The authors thank Nischal Guruwacharya from South Dakota State University for reviewing the manuscript content and Niranjan Bhujel from the University of Maine for his assistance in interface file generation.

\section*{Data Availability}
Upon paper acceptance, the MATLAB/Simulink model, Simulink-Python interface code, and partitioned modeling code will be hosted on GitHub using an open-source license.

\bibliographystyle{IEEEtran}
\bibliography{references.bib}

\section*{appendix}
\label{section:appendix}

The error measures for the quantitative assessment of the estimation are given in Table~I. The single-phase feeder adjusted line parameters are given in Table~III. Table~IV shows the single-phase feeder line parameters, and Table~V shows residential consumers' active and reactive ZIP load coefficients.

\begin{table}[!h]
\caption*{Table III \\ Single-Phase Low Voltage Transformer Parameters.}
\centering
\label{tab:transformer_parameters}
\begin{tabular}{c|c}
\hline
Transformer Parameters            & Value  \\ \hline
Primary winding resistance  & 0.06 p.u.                      \\ \hline
Primary winding leakage inductance  & 0.020 p.u.                  \\ \hline
Secondary winding resistance & 0.0264 p.u.   \\ \hline
Secondary winding leakage inductance & 0.0550 p.u.   \\ \hline
Magnetizing resistance & 500 p.u.   \\ \hline
Magnetizing inductance & 500 p.u.   \\ \hline
\end{tabular}
\end{table}

\vspace{-1em}

\begin{table}[!h]
\caption*{Table IV \\ Single-Phase Feeder Line Parameters.}
\centering
\label{tab:summary_parameters}
\begin{tabular}{c|c|c}
\hline
Parameters            & Drop-off Lines & Pole-to-pole Lines \\ \hline
Resistance ($\Omega$/\text{km}) & 0.549          & 0.346             \\ \hline
Inductance (\text{mH/km})  & 0.23           & 0.24              \\ \hline
Capacitance ($\mu$\text{F}/km)   & 0.055           & 0.072              \\ \hline
\end{tabular}
\end{table}

\begin{table}[!h]
\caption*{Table V \\ Active and Reactive ZIP Load Coefficients for Residential Customers Class of Stratum D.}
\label{tab:ZIPcoff}
\centering

\begin{tabular}{c|c|c|c|c|c|c}
\hline
\multicolumn{1}{c}{\begin{tabular}[c]{@{}c@{}}Consumer\\ Class\end{tabular}} & $Z_p$ & $I_p$ & $P_p$ & $Z_q$ & $I_q$ & $P_q$ \\ \hline
Stratum A & 1.5 & -2.31 & 1.81 & 7.41 & -11.97 & 5.55 \\ \hline
Stratum B & 1.57 & -2.48 & 1.91 & 9.28 & -15.29 & 7.01 \\ \hline
Stratum C & 1.56 & -2.49 & 1.93 & 10.1 & -16.75 & 7.65 \\ \hline
Stratum D & 1.31 & -1.94 & 1.63 & 9.20 & -15.27 & 7.07 \\ \hline
Stratum E & 0.96 & -1.17 & 1.21 & 6.28 & -10.16 & 4.88 \\ \hline
Stratum F & 1.18 & -1.64 & 1.47 & 8.29 & -13.67 & 6.38 \\ \hline
\end{tabular}
\end{table}

\begin{table*}[!h]
\centering
\renewcommand{\arraystretch}{2.5}
\vspace{-50em}
\caption*{Table I \\ Quantitative assessment of the estimation using error measures.}
\label{tab:metrics}
\begin{tabular}{>{\small}p{6.5cm}>{\small}p{4cm}>{\small}p{6.5cm}}


\hline \centering \textbf{Name} & \centering\textbf{Definition} & \textbf{Explanation}  \\ 
\hline 
\hline Adjusted $R^2$ Correlation Coefficient & \centering $1-\frac{N-1}{N-p}(\frac{\Sigma (y_i - \hat{y}_i)^2}{ \Sigma (y_i - \bar{y})^2})$ & \Centerstack{$N$: number of data points\\ $p$ = regression coefficients number\\ y : true value\\ $\bar{y}$ : mean value of y \\ $\hat{y}$ : predicted value}\\  

\hline  Normalized Root Mean Squared Error (NRMSE) &\centering  $\frac{\left \| y(t) - \hat{y}(t) \right \|_2}{\left \| y(t) - {\;}{\bar{y}(t)} \right \|_2}$ & \Centerstack{${y}$ : true value\\ $\hat{y}$ : predicted value\\ $\bar{y}$ : mean of true value} \\

\hline Corrected Akaike's Information Criterion (AICc) & \centering $-2 \mathcal{L}(\alpha) +  2d + \frac{2d(d+1)}{N-d-1}$ & \Centerstack{$\mathcal{L}$ : log-likelihood function\\ $\alpha$ : parameter indicating model complexity\\ $d$ : effective number of parameters\\ $N$: number of data points}\\

\hline Bayes information criterion (BIC) & \centering $-2 \mathcal{L}(\alpha) + dlog(N)$ & \Centerstack{$\mathcal{L}$ : log-likelihood function\\ $\alpha$ : parameter indicating model complexity\\ $d$ : effective number of parameters\\ $N$ : number of data points}\\
\hline
 \end{tabular}
\end{table*}

\end{document}